\documentclass[
  reprint,
  amsmath,
  amssymb,
  aps,
]{revtex4-2}

\usepackage[utf8]{inputenc}

\usepackage{amsfonts}
\usepackage{amsthm}
\usepackage{mathtools}
\usepackage{bm}
\usepackage{braket}

\DeclareMathOperator{\Tr}{Tr}

\usepackage{graphicx}
\usepackage{xcolor}

\usepackage{tikz-cd}
\usepackage{quiver}

\usepackage[justification=raggedright,singlelinecheck=false]{caption}
\usepackage{subcaption}

\usepackage{dcolumn}   
\usepackage{booktabs}  

\usepackage{soul}      
\usepackage{lipsum}    

\usepackage[
  colorlinks=true,
  citecolor=blue,
  urlcolor=blue,
  linkcolor=blue
]{hyperref}

\begin{document}

\preprint{APS/123-QED}

\title{State Preparation Protocols for Entangled States via Open Quantum Walks}

\author{Pedro Linck Maciel}
 \email{Contact author: pedro.linck@ufpe.br}
\author{Nadja K. Bernardes}%
\affiliation{Departamento de Física, Centro de Ciências Exatas e da Natureza, Universidade Federal de Pernambuco, Recife-PE, Brazil, 50670-901}



\date{\today}

\begin{abstract}
Open quantum walks couple transitions on a graph to quantum operations on
an internal degree of freedom. We use this structure to formulate protocols
for quantum state preparation with nonunitary Kraus operators. We construct
a ring-shaped OQW preparing an ensemble of Dicke states, with W states
appearing as the single-excitation case, from which any individual Dicke state is
recovered by postselecting the walker position; its convergence is governed by the
spectral gap of the underlying Markov chain, for which we obtain a closed-form
approximate expression. For GHZ states we present a two-node protocol using
Kraus operators built from projective measurements that prepares them
deterministically without requiring a measurement of the walker, and analyze
how unsharp measurements affect the results and the convergence of the walk.
We also show that the quantum trajectories method embeds naturally in the
OQW framework as a graph-structured collision model.

\end{abstract}

\maketitle

\section{\label{intro} Introduction}

\par An emerging view of the role of the environment in quantum computation is to use dissipation as a resource \cite{Cirac09, korkmaz20, Petr12}. This area is called dissipative quantum computation (DQC). An efficient way to implement dissipative quantum algorithms is via open quantum walks (OQWs) \cite{Petr12, Petr12-1, Dutta25}. OQWs are a type of quantum walk in which the dynamics is driven entirely by the environment. The main model of DQC using OQWs has a linear topology and all Kraus operators are proportional to unitaries \cite{Petr12, linck25}; this model can implement any convex combination of unitary maps and has well-known dynamical and thermodynamical properties \cite{linck25, linck2026}.

\par Efficient and reliable quantum state preparation is a crucial task in quantum computing; some quantum algorithms assume a given initial state is already prepared, without explicitly defining how it is prepared. Some special entangled states are useful for quantum computation and communication tasks \cite{QCI}, for example, W states \cite{Yi2023,Xing2025}, GHZ states \cite{Pickston2023, Hua2025}, and Dicke states \cite{Bai2018, Illiano2024, Roga2023, andreas19, maitra20}.

\par Quantum state preparation is the task of driving a register from a fixed
fiducial state, typically $\ket{0}^{\otimes N}$, to a prescribed target
state~\cite{QCI,Plesch2011}. It is a primitive across quantum
computation, where a prepared register is the input to essentially every
circuit model~\cite{QCI}. There are several examples across quantum communication, where
shared entangled resources such as $W$~\cite{Yi2023,Xing2025} and
GHZ~\cite{Pickston2023,Hua2025} states underpin conference key agreement
and secret sharing. Its cost is not a technicality: several quantum
algorithms are stated relative to a specific input
state and if that initial state preparation is not itself efficient, the claimed speedup can
be lost~\cite{Aaronson2015}. The same caveat governs the quantum machine
learning routines built on top of such subroutines~\cite{Biamonte2017}.
For a generic $N$-qubit pure state the cost is provably
prohibitive: counting arguments give a lower bound exponential in $N$ on
the number of two-qubit gates required~\cite{Knill1995}, and the best
known explicit circuits attain a matching $\Theta(2^{N})$ CNOT count, and
are in that sense asymptotically optimal~\cite{Shende2006,Plesch2011}.
 
\par Initial states of practical interest are, however, highly structured, and a
large body of work exploits this. Within the unitary paradigm one finds
divide-and-conquer amplitude-encoding schemes~\cite{Araujo2021} and
deterministic combinatorial constructions for permutation-symmetric
states, in particular the explicit Dicke-state circuits of B\"artschi and
Eidenbenz~\cite{andreas19} and their compilations for near-term
hardware~\cite{maitra20}. A complementary paradigm, and the one adopted
in this work, replaces coherent control by engineered dissipation: rather
than steering the register with unitaries, one designs an environment
whose steady state is the target. This reservoir-engineering
programme was introduced by Poyatos, Cirac and Zoller~\cite{Poyatos1996},
formulated for entangled-state preparation via quantum Markov processes
by Kraus \emph{et al.}~\cite{Kraus2008}, and shown by Verstraete, Wolf
and Cirac to support universal quantum computation without any coherent
dynamics~\cite{Cirac09}. It has been realized experimentally in
open-system trapped-ion simulators~\cite{Barreiro2011}, in the
dissipative stabilization of entanglement between two trapped-ion
qubits~\cite{Lin2013}, and in steady-state entanglement of two
macroscopic atomic ensembles~\cite{Krauter2011}. Dissipative schemes
targeting precisely the states considered here have also been proposed,
including the preparation of large $W$ states in optical
cavities~\cite{Sweke2013}.
 
\par Open quantum walks sit naturally at the interface between these
paradigms. They realize a family of completely positive, trace-preserving
maps on a graph-structured register~\cite{Petr12-1}, they provide an
efficient route to dissipative quantum computation~\cite{Petr12,linck25},
and, as we show below, they give a unified and physically transparent
language for preparing GHZ, $W$ and Dicke states. In the constructions
presented here the graph degree of freedom simultaneously drives the
dynamics and retains an explicit classical record of the measurement
outcomes, so that the prepared state is identified by the walker position
rather than inferred.

\par The main contribution of this paper is to show that open quantum walks can provide a useful tool for dissipative quantum computation using more general Kraus operators and graph topologies, providing a framework for preparing GHZ, W, and Dicke states, and to analyze, both analytically and numerically, the convergence rate and optimal parameters for the computation. We also provide a natural embedding of the quantum trajectories method to simulate the continuous evolution of open quantum systems in the OQW framework.
\par The paper is organized as follows. In Sec.~\ref{sec:oqw}, we introduce the basic
definitions and results needed to follow our work. In Sec.~\ref{sec:ghzstab}, we
present a family of protocols for preparing GHZ states via OQWs and show how the
graph degree of freedom can be used to stabilize the target against phase errors, and to
accommodate unsharp measurements. In Sec.~\ref{sec:stateprep}, we present a
protocol for preparing Dicke and $W$ states in the OQW framework, analyze its
steady state and convergence analytically and numerically, obtaining a
closed-form approximate expression for the spectral gap that governs the convergence time,
and discuss the locality and the physical cost of the Kraus operators involved.
In Sec.~\ref{sec:quantum-traj}, we provide a canonical embedding of the quantum
trajectories method into the OQW formalism, and show that the resulting
construction relates structurally to a collision model. In
Sec.~\ref{sec:conclusion}, we conclude with a brief summary of our results and
outline perspectives for future research.

\section{\label{sec:oqw}Open Quantum Walks } 

\par Quantum walks provide models of quantum dynamics on a graph. The vertices label an orthonormal basis of the position Hilbert space, and the walker moves between vertices according to the graph connectivity together with its internal degrees of freedom~\cite{QWSA}. In discrete time, the evolution is obtained by iterating a prescribed one-step update. In this work we concentrate on open quantum walks, a class of quantum walks in which the effective dynamics is entirely generated by interaction with an environment.

\par Any quantum evolution can be described by a completely positive and trace-preserving map $\Lambda\colon \mathcal{B}(\mathcal{V}) \to \mathcal{B}(\mathcal{V})$, where $\mathcal{B}(\mathcal{V})$ denotes the algebra of bounded operators acting on the Hilbert space $\mathcal{V}$~\cite{OQS}. It is well known~\cite{OQS} that if $\dim(\mathcal{V})=d$, then $\Lambda$ admits an operator-sum (Kraus) decomposition, $\Lambda(\rho)=\sum_{i=0}^{m-1} K_i\,\rho\,K_i^\dagger$ with $m \leq d^2$, where the Kraus operators $\{K_i\}_{i=0}^{m-1}$ satisfy the completeness condition $\sum_{i=0}^{m-1} K_i^\dagger K_i = I$. This decomposition is not unique.

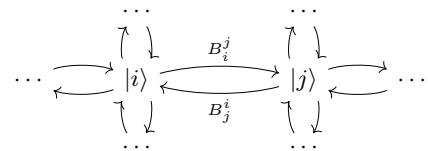
\begin{figure}[b]
\[\begin{tikzcd}[row sep=scriptsize]
	& \cdots && \cdots \\
	\cdots & {\ket{i}} && {\ket{j}} & \cdots \\
	& \cdots && \cdots
	\arrow[curve={height=-6pt}, from=1-2, to=2-2]
	\arrow[curve={height=-6pt}, from=1-4, to=2-4]
	\arrow[curve={height=-6pt}, from=2-1, to=2-2]
	\arrow[curve={height=-6pt}, from=2-2, to=1-2]
	\arrow[curve={height=-6pt}, from=2-2, to=2-1]
	\arrow["{B_i^j}", curve={height=-6pt}, from=2-2, to=2-4]
	\arrow[curve={height=-6pt}, from=2-2, to=3-2]
	\arrow[curve={height=-6pt}, from=2-4, to=1-4]
	\arrow["{B_j^i}", curve={height=-6pt}, from=2-4, to=2-2]
	\arrow[curve={height=-6pt}, from=2-4, to=2-5]
	\arrow[curve={height=-6pt}, from=2-4, to=3-4]
	\arrow[curve={height=-6pt}, from=2-5, to=2-4]
	\arrow[curve={height=-6pt}, from=3-2, to=2-2]
	\arrow[curve={height=-6pt}, from=3-4, to=2-4]
\end{tikzcd}\]\caption{\label{fig:OQW} An arbitrary open quantum walk can be represented by this visual diagram. If there is an omitted edge in a particular diagram, this means that the corresponding operator $B_i^j$ is zero. Figure extracted from Ref.~\cite{linck25}.}
\end{figure}

\par To define an open quantum walk, we consider an internal (coin) Hilbert space $\mathcal{H}$ for the walker and a position (graph) Hilbert space $\mathcal{G}$ on which the walk is defined. Let $\{\ket{i}\}_{i\in G}$ be an orthonormal basis of $\mathcal{G}$, where $G$ denotes the vertex set of the underlying graph. For each $i\in G$, we assign a family of linear operators $\{B_i^{\,j}\}_{j\in G}$, with $B_i^{\,j}\colon \mathcal{H}\to\mathcal{H}$, which act on the internal state when a transition from vertex $i$ to vertex $j$ occurs. These operators satisfy the completeness condition
\begin{equation}
\label{eqn:kraus_cond_oqw}
    \sum_{j\in G} B_i^{\,j\dagger} B_i^{\,j} = I,
\end{equation}
which guarantees trace preservation of the resulting dynamics. To encode the jump together with the corresponding transformation of the internal state, we introduce the operators $M_i^{\,j}=B_i^{\,j}\otimes \ket{j}\!\bra{i}\in \mathcal{B}(\mathcal{H}\otimes \mathcal{G})$. One readily checks that Eq.~\ref{eqn:kraus_cond_oqw} implies the completeness relation $\sum_{i,j \in G} M_i^{\,j\dagger} M_i^{\,j} = I$, so that the collection $\{M_i^{\,j}\}_{i,j\in G}$ defines a quantum channel. This operator-sum description specifies an open quantum walk. Pictorially, an OQW can be represented by a directed graph in which the edge $i\to j$ is labeled by the operator $B_i^{\,j}$, as illustrated in Fig.~\ref{fig:OQW}.

Given an initial state $\rho_0\in \mathcal{B}(\mathcal{H}\otimes \mathcal{G})$, the evolution is defined iteratively by
\begin{equation}
\label{eqn:recursive_oqw}
\begin{cases}
\rho^{[0]}=\rho_0,\\
\rho^{[n]}=\displaystyle\sum_{i,j \in G} M_i^{\,j}\,\rho^{[n-1]}\,M_i^{\,j\dagger}, \qquad n\ge 1,
\end{cases}
\end{equation}
where $\rho^{[n]}$ denotes the open-quantum-walk state after $n$ steps. It is shown in Ref.~\cite{Petr12-1} that if the initial state admits the decomposition $\rho_0=\sum_{i,j\in G}\rho_{ij}\otimes \ket{j}\!\bra{i}$,with trace-class operators $\rho_{ij}$ satisfying $\sum_{i\in G}\Tr(\rho_{ii})=1$, then for every $n\ge 1$ the state $\rho^{[n]}$ becomes block diagonal in the graph basis, $\rho^{[n]}=\sum_{i\in G}\rho_{ii}^{[n]}\otimes \ket{i}\!\bra{i}$,where the trace-class operators $\rho_{ii}^{[n]}$ are generated by the iterative update in Eq.~\ref{eqn:recursive_oqw}. In particular, a single step is sufficient to eliminate coherences between different vertices.

\begin{figure}[h!]
\[\begin{tikzcd}
	{\ket{0}} && {\ket{1}} && \cdots && {\ket{N-1}}
	\arrow["{\sqrt{\lambda} I}", from=1-1, to=1-1, loop, in=55, out=125, distance=10mm]
	\arrow["{\sqrt{\omega} U_0}", curve={height=-6pt}, from=1-1, to=1-3]
	\arrow["{\sqrt{\lambda} U_0^\dagger}", curve={height=-6pt}, from=1-3, to=1-1]
	\arrow["{\sqrt{\omega} U_1}", curve={height=-6pt}, from=1-3, to=1-5]
	\arrow["{\sqrt{\lambda} U_1^\dagger}", curve={height=-6pt}, from=1-5, to=1-3]
	\arrow["{\sqrt{\omega} U_{N-2}}", curve={height=-6pt}, from=1-5, to=1-7]
	\arrow["{\sqrt{\lambda} U_{N-2}^\dagger}", curve={height=-6pt}, from=1-7, to=1-5]
	\arrow["{\sqrt{\omega} I}", from=1-7, to=1-7, loop, in=55, out=125, distance=10mm]
\end{tikzcd}\]
\caption{\label{fig:linearOQW} The diagram corresponding to the linear OQW model. Each $U_i$ is a unitary operator, and $\omega, \lambda \geq 0$ are such that $\omega + \lambda = 1$. Figure extracted from Ref.~\cite{linck25}.}
\end{figure}
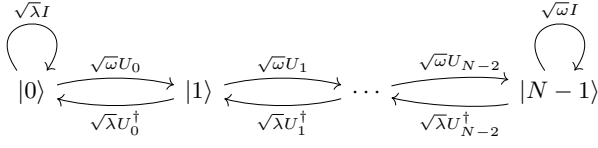

It is also convenient to note that the diagrammatic description in Fig.~\ref{fig:OQW}, together with an initial state, uniquely fixes the dynamics. For a given graph, if the directed edge $\ket{i}\to\ket{j}$ carries the label $B_i^j$, then this label is incorporated into the update rule in Eq.~\ref{eqn:recursive_oqw} by setting
$M_i^j = B_i^j \otimes \ket{j}\bra{i}$.
Therefore, once all edges and their associated operators are specified (with absent edges understood to correspond to the zero Kraus operator), the full collection $\{M_i^j\}$ can be constructed explicitly. As an example, for the linear walk in Fig.~\ref{fig:linearOQW} we have
$M_0^0 = \sqrt{\lambda}\, I \otimes \ket{0}\bra{0}$,
$M_0^1 = \sqrt{\omega}\, U_0 \otimes \ket{1}\bra{0}$,
$M_1^0 = \sqrt{\lambda}\, U_0^\dagger \otimes \ket{0}\bra{1}$,
and $M_1^1 = 0$ (for $N>2$ there is no directed edge from $\ket{1}$ to $\ket{1}$), and similarly for the remaining nodes, for a fixed finite set of unitaries $\{U_i\}$ and nonnegative parameters $\omega,\lambda$ satisfying $\omega+\lambda=1$.

\section{Protocols for preparing GHZ states via open quantum walks}
\label{sec:ghzstab}

In this section we present a family of OQW protocols for preparing GHZ states,
beginning with a measurement-conditioned scheme and showing how the graph degree
of freedom can be used to convert it into a relaxation scheme, to stabilize the
target against noise, and to accommodate imperfect measurements.

\subsection{Measurement-conditioned preparation}
\label{sec:ghzcond}

\begin{figure}[b]
\[
\begin{tikzcd}
{\ket{0}} && {\ket{1}}
\arrow["{K_+}", from=1-1, to=1-1,
loop, in=60, out=120, distance=5mm]
\arrow["{K_-}", curve={height=-6pt}, from=1-1, to=1-3]
\arrow["{K_-}", curve={height=-6pt}, from=1-3, to=1-1]
\arrow["{K_+}", from=1-3, to=1-3,
loop, in=60, out=120, distance=5mm]
\end{tikzcd}
\]
\caption{\label{fig:ghz_oqw}
Two-node OQW for preparing GHZ-type states. The operator $K_+$ keeps the walker
at the same node, while $K_-$ transfers it to the other node.}
\end{figure}
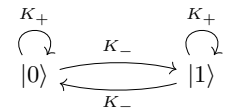

We consider an OQW with an internal Hilbert space of $N$ qubits and a graph space of $1$ qubit. On the internal space, define
\begin{equation}
K_\pm=\frac{I\pm X^{\otimes N}}{2},
\label{eq:Kpm}
\end{equation}
the projectors onto the $\pm1$ eigenspaces of $X^{\otimes N}$, so that
\begin{equation}
K_\pm^\dagger K_\pm=K_\pm,
\qquad
K_+K_-=0,
\qquad
K_+ + K_-= I.
\label{eq:Kproj}
\end{equation}
Therefore $\{K_+,K_-\}$ is a valid set of Kraus operators, corresponding
physically to a projective measurement of the global stabilizer $X^{\otimes N}$.
Following the convention of Sec.~\ref{sec:oqw}, in which $B_i^{\,j}$ denotes the
transition operator from node $i$ to node $j$, we define the two-node OQW
\begin{equation}
  \begin{aligned}
    B_{0}^{\,0} &= K_{+}, &\qquad B_{0}^{\,1} &= K_{-}, \\
    B_{1}^{\,0} &= K_{-}, &\qquad B_{1}^{\,1} &= K_{+} ,
  \end{aligned}
  \label{eq:ghzB}
\end{equation}
depicted in Fig.~\ref{fig:ghz_oqw}. Thus $K_+$ keeps the walker at the same node
and $K_-$ transfers it to the other node, and by Eq.~\eqref{eq:Kproj} the
normalization condition $\sum_j B_i^{\,j\dagger}B_i^{\,j}= I$ holds for
$i=0,1$.

We initialize the OQW as $\rho^{(0)}=\ket{0^N}\bra{0^N}\otimes\ket{0}\bra{0}$
and write $\rho^{(n)}=\sum_{j} \rho_j^{(n)}\otimes\ket{j}\bra{j}$, with
$\rho_j^{(n+1)}=\sum_i B_i^{\,j}\rho_i^{(n)}B_i^{\,j\dagger}$. Defining
\begin{equation}
\ket{\mathrm{GHZ}_\pm}=\frac{\ket{0^N}\pm\ket{1^N}}{\sqrt{2}},
\label{eq:ghzdef}
\end{equation}
we have $K_\pm\ket{0^N}=\tfrac{1}{\sqrt2}\ket{\mathrm{GHZ}_\pm}$, so the first
step gives
\begin{equation}
\rho^{(1)}
=
\frac{1}{2}\ket{\mathrm{GHZ}_+}\bra{\mathrm{GHZ}_+}\otimes\ket{0}\bra{0}
+
\frac{1}{2}\ket{\mathrm{GHZ}_-}\bra{\mathrm{GHZ}_-}\otimes\ket{1}\bra{1}.
\label{eq:rho1}
\end{equation}
Measuring the graph position therefore prepares a GHZ-type internal state: outcome
$\ket{0}$ gives $\ket{\mathrm{GHZ}_+}$ and outcome $\ket{1}$ gives
$\ket{\mathrm{GHZ}_-}$, the latter converted into the former by a Pauli $Z$ on
one internal qubit.

Since $K_+\ket{\mathrm{GHZ}_+}=\ket{\mathrm{GHZ}_+}$,
$K_-\ket{\mathrm{GHZ}_-}=\ket{\mathrm{GHZ}_-}$ and
$K_\mp\ket{\mathrm{GHZ}_\pm}=0$, the $\ket{\mathrm{GHZ}_+}$ part of the walker state remains
stationary in a fixed node while the $\ket{\mathrm{GHZ}_-}$ part changes node at each step.
In the second step we have
\begin{equation}
\rho^{(2)}
=
\left(
\frac{1}{2}\ket{0^N}\bra{0^N}+\frac{1}{2}\ket{1^N}\bra{1^N}
\right)\otimes\ket{0}\bra{0},
\label{eq:rho2}
\end{equation}
and $\rho^{(3)}=\rho^{(1)}$, so that the dynamics is periodic,
\begin{equation}
\rho^{(2m+1)}=\rho^{(1)},
\qquad
\rho^{(2m)}=\rho^{(2)},
\qquad
m\geq 1.
\label{eq:periodic}
\end{equation}
This protocol is not a relaxation scheme toward a unique stationary state, but a
measurement-conditioned one: at every odd step, measuring the walker prepares
either $\ket{\mathrm{GHZ}_+}$ or $\ket{\mathrm{GHZ}_-}$, and a conditional
single-qubit phase correction converts both outcomes into $\ket{\mathrm{GHZ}_+}$.

\subsection{Correction on the edge}
\label{sec:edgecorr}

Let $Z_1$ be the Pauli $Z$ operator acting on the first internal qubit. Since
$Z_1$ anticommutes with $X^{\otimes N}$ we have $Z_1 K_\pm = K_\mp Z_1$, and in
particular
\begin{equation}
  Z_1 \ket{\mathrm{GHZ}_-} = \ket{\mathrm{GHZ}_+}.
  \label{eq:Zcorr}
\end{equation}
Instead of applying this correction as a classical feedforward operation
conditioned on a measurement of the walker, we incorporate it into the transition
operator of the edge $\ket{1}\to\ket{0}$, defining
\begin{equation}
  \begin{aligned}
    B_0^{\,0} &= K_+, &\qquad B_0^{\,1} &= K_-, \\
    B_1^{\,0} &= Z_1 K_-, &\qquad B_1^{\,1} &= Z_1 K_+ .
  \end{aligned}
  \label{eq:corrB}
\end{equation}
Because $Z_1$ is unitary, $\sum_j B_i^{\,j\dagger}B_i^{\,j}=K_++K_-= I$,
so Eq.~\eqref{eq:corrB} defines a valid OQW.

The first step again reproduces Eq.~\eqref{eq:rho1}. At the second step we have
\begin{equation}
  \rho^{(2)} = \ket{\mathrm{GHZ}_+}\bra{\mathrm{GHZ}_+} \otimes \ket{0}\bra{0},
  \label{eq:absorb}
\end{equation}
with $\rho^{(n)}=\rho^{(2)}$ for all $n\geq2$. The state \eqref{eq:absorb} is
a steady-state. No measurement of the walker is performed, no correction is applied by
hand, and the preparation succeeds with unit probability in exactly two steps.
Adding the $Z_1$ correction to $B_1^0$ has converted the periodicity of
Eq.~\eqref{eq:periodic} into convergence.

Since $K_\pm$ project onto the $\pm1$ eigenspaces of $X^{\otimes N}$, each of
dimension $2^{N-1}$, rather than onto the GHZ states themselves,
Eq.~\eqref{eq:absorb} is the unique attractor only within the invariant manifold
\begin{equation}
  \mathcal{M} = \mathrm{span}\bigl\{ \ket{0^N},\, \ket{1^N} \bigr\},
  \label{eq:manifold}
\end{equation}
which contains the initial state. On the full internal Hilbert space, every state
supported in the $+1$ eigenspace of $X^{\otimes N}$ at node $\ket{0}$ is
stationary. Lifting this restriction requires measuring the entire stabilizer
group.

\subsection{Unsharp measurement operators}
\label{sec:unsharp}

The projectors $K_\pm$ describe an ideal projective measurement of
$X^{\otimes N}$. A more realistic model replaces them by the unsharp measurement
operators
\begin{equation}
  \begin{aligned}
    M_\pm &= \sqrt{\frac{I \pm \eta\, X^{\otimes N}}{2}}
           = c\,K_\pm + s\,K_\mp , \\[2pt]
    c &= \sqrt{\frac{1+\eta}{2}},
    \qquad
    s = \sqrt{\frac{1-\eta}{2}},
  \end{aligned}
  \label{eq:unsharp}
\end{equation}
where $\eta\in[0,1]$ is the sharpness: $\eta=1$ recovers the projectors $K_\pm$,
while $\eta=0$ gives $M_\pm=I/\sqrt{2}$, a measurement that returns no
information. Since $M_+^\dagger M_+ + M_-^\dagger M_- =(c^2+s^2)(K_++K_-)= I$,
the pair is a valid POVM for every $\eta$.

The walk then possesses a unique stationary state $\rho^{\,\mathrm{ss}}$ on
$\mathcal{M}\otimes\mathcal{G}$ (for $\eta>0$), block diagonal in the graph basis
with node-resolved components
$\rho_i^{\,\mathrm{ss}}=\bra{i}\rho^{\,\mathrm{ss}}\ket{i}$, each diagonal in the
GHZ basis,
\begin{equation}
  \begin{aligned}
    \rho_0^{\,\mathrm{ss}} &= c^4 \ket{\mathrm{GHZ}_+}\bra{\mathrm{GHZ}_+}
                            + s^4 \ket{\mathrm{GHZ}_-}\bra{\mathrm{GHZ}_-}, \\
    \rho_1^{\,\mathrm{ss}} &= c^2 s^2 \bigl(
       \ket{\mathrm{GHZ}_+}\bra{\mathrm{GHZ}_+}
     + \ket{\mathrm{GHZ}_-}\bra{\mathrm{GHZ}_-} \bigr),
  \end{aligned}
  \label{eq:ssunsharp}
\end{equation}
normalized as $\mathrm{Tr}\,\rho_0^{\,\mathrm{ss}}+\mathrm{Tr}\,\rho_1^{\,\mathrm{ss}}
=(c^2+s^2)^2=1$. The unconditional fidelity with the target is
\begin{equation}
  F = \bra{\mathrm{GHZ}_+} \mathrm{Tr}_{\mathcal{G}}\, \rho^{\,\mathrm{ss}}
      \ket{\mathrm{GHZ}_+}
    = c^2 = \frac{1+\eta}{2},
  \label{eq:Funcond}
\end{equation}
degrading linearly with the sharpness and reaching the trivial value $1/2$ at
$\eta=0$. Measuring the graph register and postselecting the outcome $\ket{0}$,
however, yields
\begin{equation}
  \begin{aligned}
  p_0 = \mathrm{Tr}\,\rho_0^{\,\mathrm{ss}} = c^4+s^4 = \frac{1 + \eta^2}{2}\\
  F_0 = \frac{c^4}{c^4 + s^4} = \frac{(1+\eta)^2}{2\,(1+\eta^2)} ,
  \label{eq:Fcond}
  \end{aligned}
\end{equation}
with
\begin{equation}
  F_0 - F = \frac{\eta\,(1-\eta)(1+\eta)}{2\,(1+\eta^2)} > 0
  \qquad (0<\eta<1).
  \label{eq:advantage}
\end{equation}
An imperfect measurement of the internal register is thus partially compensated
by a projective measurement of the walker, a single qubit. The gain $F_0-F$
vanishes at both endpoints and is maximal at $\eta^\star=\sqrt{\sqrt{5}-2}\approx0.49$.

Denote by $\Lambda_\eta$ the map defined by Eq.~\eqref{eq:corrB} with $K_\pm$
replaced by $M_\pm$, restricted to $\mathcal{B}(\mathcal{M}\otimes\mathcal{G})$.
Its spectrum consists of the stationary eigenvalue $1$, the eigenvalue
$\lambda_2(\eta)=\sqrt{1-\eta^2}=2cs$, and is zero otherwise, so the asymptotic
convergence is set by the spectral gap
\begin{equation}
  \Delta(\eta) = 1 - \lambda_2(\eta) = 1 - \sqrt{1-\eta^2}
  \simeq \frac{\eta^2}{2} \quad (\eta \ll 1).
  \label{eq:gapeta}
\end{equation}
A weak measurement ($\eta\ll1$) therefore relaxes the walk only at a rate
quadratic in the sharpness, with
$t_{\mathrm{conv}}\simeq\ln(1/\varepsilon)/\Delta(\eta)$; the gap closes as
$\eta\to0$, consistent with the loss of a unique stationary state there.

At $\eta=1$ the eigenvalue $\lambda_2$ coalesces with the zero eigenvalues, the
corresponding eigenvectors merge, and $\Lambda_1$ ceases to be diagonalizable:
it possesses a Jordan block of size two at $\lambda=0$. The projective limit is
thus a defective (exceptional) point of the family $\{\Lambda_\eta\}$, at which
the exponential relaxation of Eq.~\eqref{eq:gapeta} is replaced by exact
termination in two steps.

\section{\label{sec:stateprep}Dicke state preparation via Open Quantum Walks}
In this section, we explore how to prepare an ensemble of Dicke states with weights that can be tuned according to environmental parameters.

\subsection{Dicke states}
Given a Hilbert space $\mathcal{H}$ of $N$ qubits, the Dicke state $\ket{\mathcal{D}^N_k}$ is defined as an equal-weight superposition of all $\binom{N}{k}$ distinct permutations of the state $\ket{1\dots 1\,0\dots 0}$ with exactly $k$ qubits in the state $\ket{1}$. Mathematically this can be written as 
\begin{equation}
\ket{\mathcal{D}^N_k} = \binom{N}{k}^{-1/2} \sum_{x: |x| = k} \ket{x_1\dots x_N},
\end{equation}
where $x_i$ are the components of a bit string $x \in \{0,1\}^N$, and $|x|$ its Hamming weight (i.e., the number of $1$'s in the string). We define the jump operator acting on one qubit as
\begin{equation}
    J = \sqrt{p}\,\ket{0}\bra{1},
\end{equation}
that can be interpreted as the spontaneous decay from the excited state to the ground state. The Dicke problem addresses collective superradiant decay \cite{rosario25}, where the master equation has a collective-jump-operator Lindbladian. Since we focus on discrete evolutions, we can discretize it using well-known methods (see Eqs.~\ref{eqn:krauszero} and \ref{eqn:krausops}), together with a slight modification of the referred Lindbladian, to define a quantum channel consisting of the Kraus operators
\begin{equation}
    J_c = \dfrac{1}{A}\left(J_1+\cdots + J_N\right), \qquad K_c = \sqrt{I - J_c^\dagger J_c},
\end{equation}
where $A$ is a tunable parameter chosen such that $I-J_c^\dagger J_c$ is positive semidefinite, and $J_i$ is the jump operator $J$ acting on the $i$th qubit. We will need the relations
\begin{equation}
\label{eqn:dicke_sink}
J_c^m\ket{\mathcal{D}^N_k} = 0, \qquad m > N,
\end{equation}
\begin{equation}
\label{eqn:jump_dicke}
J_c\ket{\mathcal{D}_k^N} = \sqrt{\frac{k p (N-k+1)}{A^2}}\, \ket{\mathcal{D}_{k-1}^N},
\end{equation}
\begin{equation}
\label{eqn:staystill_dicke}
K_c \ket{\mathcal{D}_k^N} = \sqrt{1 - \frac{k p (N-k+1)}{A^2}}\, \ket{\mathcal{D}_k^N},
\end{equation}
which are proven in Appendix~\ref{app:dicke-relations}. Therefore, starting from $\ket{1}^{\otimes N}$ and applying the operator $J_c$ iteratively $k$ times, we reach the Dicke state $\ket{\mathcal{D}^N_{N-k}}$. Note that, to ensure $1 - k p (N-k+1)/A^{2} \ge 0$ for all
$0 \le k \le N$, we must require $A^{2} \ge p \max_{k} k(N-k+1)$.
The maximum over integer $k$ is attained at $k = \lceil (N+1)/2 \rceil$
and equals
\begin{equation}
  \max_{0 \le k \le N} k(N-k+1)  = \Big\lfloor \frac{(N+1)^{2}}{4} \Big\rfloor 
  \label{eq:maxk}
\end{equation}
so that the exact positivity condition reads
\begin{equation}
  A \;\ge\; \sqrt{\,p \Big\lfloor \tfrac{(N+1)^{2}}{4} \Big\rfloor\,}.
  \label{eq:Aexact}
\end{equation}
In particular the simpler sufficient condition
\begin{equation}
  A \;\ge\; \frac{\sqrt{p}}{2}\,(N+1)
  \label{eq:Asuff}
\end{equation}
always holds, and it is tight for odd $N$; for even $N$ it overestimates
the required $A^{2}$ by $p/4$. Thus the error committed in replacing
$\max_{k} k(N-k+1)$ by $(N+1)^{2}/4$ is at most $1/4$, and vanishes for
odd $N$.

\subsection{\label{sec:dicke_oqw}Preparing Dicke states}
\begin{figure}[b]
\[\begin{tikzcd}
	{\ket{0}} && {\ket{1}} \\
	\\
	& {\ket{2}}
	\arrow["{K_c}", from=1-1, to=1-1, loop, in=60, out=120, distance=5mm]
	\arrow["{J_c}", curve={height=-6pt}, from=1-1, to=1-3]
	\arrow["{K_c}", from=1-3, to=1-3, loop, in=60, out=120, distance=5mm]
	\arrow["{J_c}", curve={height=-6pt}, from=1-3, to=3-2]
	\arrow["{(X\otimes \cdots \otimes X)J_c}", curve={height=-6pt}, from=3-2, to=1-1]
	\arrow["{K_c}"', from=3-2, to=3-2, loop, in=300, out=240, distance=5mm]
\end{tikzcd}\]
\caption{\label{fig:dicke_oqw_firstproposal} OQW model to implement Dicke states for $N = 3$.}
\end{figure}
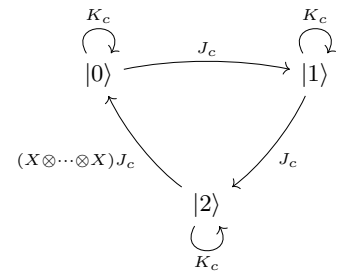

We first propose the OQW defined formally as
\begin{equation}
    M_i^{i+1} = J_c \otimes \ket{i+1} \bra{i}, \; i < N-1,
\end{equation}
\begin{equation}
    M_{N-1}^0 = [(X\otimes \cdots \otimes X)J_c] \otimes \ket{0} \bra{N-1},
\end{equation}
\begin{equation}
    M_i^i = K_c \otimes \ket{i}\bra{i},
\end{equation}
and all other $M_i^j$ are zero. The initial state for our proposal is 
\begin{equation}
    \rho_0  = \ket{1^N}\bra{1^N} \otimes \ket{0}\bra{0}.
\end{equation}
This is an OQW with a ring topology, depicted in Figure \ref{fig:dicke_oqw_firstproposal}.
From Equations (\ref{eqn:jump_dicke}), (\ref{eqn:staystill_dicke}) we know that since our initial state is a Dicke state in the internal Hilbert space, the state after $m$ steps takes the form
\begin{equation}
\rho^{[m]} = \sum_{k=0}^{N-1} p^{(m)}
_k\ket{\mathcal{D}_{N-k}^N}\bra{\mathcal{D}_{N-k}^N} \otimes \ket{k}\bra{k},
\end{equation}
where $p_k^{(m)}$ is the probability of encountering the walker in node $\ket{k}$ after $m$ steps. Note that this state is an ensemble of Dicke states, each exhibiting a non-zero detection probability if $m \geq N$. We can post-select any desired Dicke state by measuring the graph space. This OQW formulation, together with this particular initial state, enables a precise determination of the Dicke state obtained post-measurement, as the relevant information is encoded in the graph structure. Also note that the $W$ state is the single-excitation Dicke state,
\begin{equation}
\ket{W_N}=\ket{\mathcal D^N_1}.
\end{equation}
Therefore, the same protocol also prepares $W$ states as a particular case. Note that this choice of OQW with this particular state presents an underlying classical Markov chain, so we can study its dynamics and steady state in this framework \cite{linck25}. The steady state can be calculated as 
\begin{equation}
\label{eqn:tuned_steady_state}
\pi_k = \dfrac{1/(N-k)(k+1)}{2H_{N}/(N+1)}, \; 0 \leq k \leq N-1,
\end{equation}
where $H_N = \sum_{l=1}^N 1/l \approx \log(N)$ and $\pi_k$ is the steady-state probability of being at node $\ket{k}$. Another important feature of this steady state is that it is surprisingly independent of the parameters $p$ and $A$, implying that the long-term behavior of this choice of OQW and initial state is the same regardless of the rates of the single-particle and collective decay processes. However, the entire dynamics of this OQW is highly dependent of both parameters. For example, the drift velocity is
\begin{equation}
    v(k) = p_k = \dfrac{p}{A^2}(N-k)(k+1),
\end{equation}
and therefore the mean drift velocity can be calculated as 
\begin{equation}
    v_{\text{mean}} = \frac{p(N+1)(N+2)}{6A^2},
\end{equation}
showing that it grows quadratically with the size $N$ of the graph.  
\begin{figure}[]
\includegraphics[width = 8.3cm]{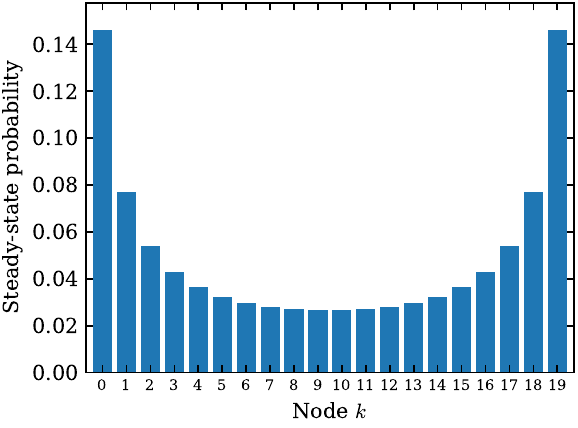}
\caption{\label{fig:steady_state_Dicke} Plot of the steady-state probability in the preparation of Dicke states as a function of the position in the graph for $N = 20$.}
\end{figure}
Our protocol prepares an ensemble, and a specific Dicke state can be
obtained by postselecting the walker at the corresponding node. It is therefore
essential to quantify the associated success probability. At long times the
walker is distributed according to Eq.~\eqref{eqn:tuned_steady_state}, so the probability of
recovering $\ket{D^N_{N-k}}$ in a single run is $\pi_k$, and the expected number
of repetitions required is
\begin{equation}
  R_k \;=\; \frac{1}{\pi_k} \;=\; \frac{2H_N\,(N-k)(k+1)}{N+1}.
  \label{eq:repetitions}
\end{equation}
The extremal states $\ket{D^N_1}$ and $\ket{D^N_N}$ are the cheapest,
$R \simeq 2H_N$, whereas the balanced states $k \simeq N/2$ are the most expensive:
\begin{equation}
  \pi_{k \simeq N/2} \;\simeq\; \frac{2}{N H_N},
  \qquad
  R_{k \simeq N/2} \;\simeq\; \frac{N H_N}{2} \;=\; \Theta(N \log N).
  \label{eq:midcost}
\end{equation}
In particular the $W$ state, $\ket{W_N} = \ket{D^N_1}$, is obtained at
node $k = N-1$ with only logarithmic overhead.

We stress that this places the present construction on a different footing from
deterministic circuit-based preparations. The Dicke-state circuits of
Ref.~\cite{andreas19} produce any $\ket{D^N_k}$ deterministically with
$O(N)$ depth, and Ref.~\cite{maitra20} provides explicit compilations for
near-term hardware. The OQW protocol proposed here is probabilistic, requires
$\Theta(N \log N)$ repetitions for balanced $k$, and employs Kraus operators that are not local. It therefore can not be compared with the most resource-efficient state-preparation protocols from unitary quantum computation. Its interest is structural: it exhibits Dicke-state
preparation as the natural stationary behaviour of a dissipative walk whose
graph register simultaneously drives the collective decay and labels the
resulting excitation sector, thereby connecting reservoir-engineering ideas to
graph-based models of dissipative computation.

Now we determine the expected number of steps $n_{1}$ required for
the walker to complete one cycle of the ring, i.e.\ to return to node
$\ket{0}$ and thereby reset the internal register to
$\ket{D^{N}_{N}}$. Since the walk is a directed cycle in which the
walker leaves node $k$ with probability $p_{k}$ at each step, the
occupation time of node $k$ is geometric with mean $1/p_{k}$, and
therefore
\begin{equation}
  n_{1} \;=\; \sum_{k=0}^{N-1} \frac{1}{p_{k}}
  \;=\; \frac{2 A^{2} H_{N}}{p\,(N+1)},
  \label{eq:n1exact}
\end{equation}
where the second equality is exactly the sum evaluated in
Eq.~\eqref{eq:sumB2} of Appendix~\ref{app:steady}. Equivalently, using
the continuum approximation,
\begin{equation}
\begin{split}
  n_{1} & \simeq\int_{0}^{N-1} \frac{dk}{v(k)} \\
  & = \frac{A^{2}}{p}\,\frac{1}{N+1}
        \int_{0}^{N-1}\!\Big(\frac{1}{N-k} + \frac{1}{k+1}\Big)\,dk \\
 & = \frac{2 A^{2}}{p\,(N+1)}\,\ln N,
  \label{eq:n1int}
\end{split}
\end{equation}
which is Eq.~\eqref{eq:n1exact} with $H_{N} \to \ln N$. Combining
Eq.~\eqref{eq:n1exact} with the positivity condition
$A^{2} \ge p (N+1)^{2}/4$ of Eq.~\eqref{eq:Asuff} gives the lower bound
\begin{equation}
  n_{1} \;\ge\; \frac{(N+1)\,H_{N}}{2}
        \;\ge\; \frac{N+1}{2}\,\ln N .
  \label{eq:n1bound}
\end{equation}

\subsection{\label{subsec:convergence-time} Convergence time for preparation of Dicke states}
\begin{figure}[htbp]
  \centering
  \begin{subfigure}{\linewidth}
    \centering
    \includegraphics[width=\linewidth]{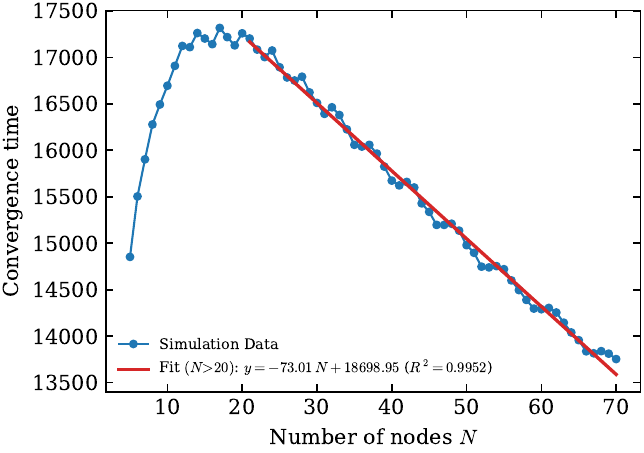}
    \caption{Convergence time in the total variation metric as a function of the number $N$ of nodes in the OQW for $A = 100$, $p = 0.9$.}
    \label{fig:metrics_0_tv}
  \end{subfigure}

  \medskip

  \begin{subfigure}{\linewidth}
    \centering
    \includegraphics[width=\linewidth]{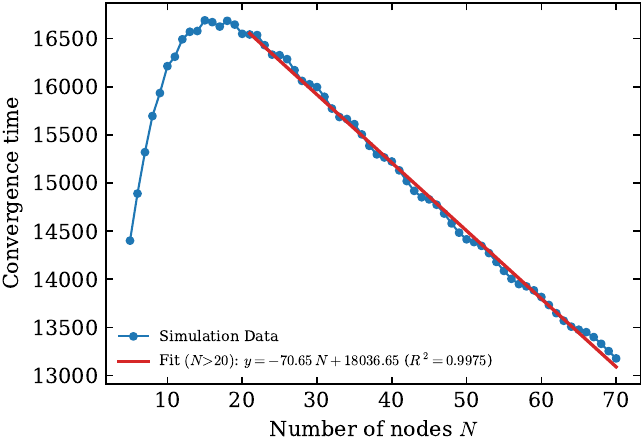}
    \caption{Convergence time in the Hellinger metric as a function of the number $N$ of nodes in the OQW for $A = 100$, $p = 0.9$.}
    \label{fig:metrics_0_hell}
  \end{subfigure}

  \caption{Total variation and Hellinger metrics to determine the convergence time for $A = 100$, $p = 0.9$.}
  \label{fig:metrics_0}
\end{figure}

The total variation distance between two probability distributions $P,Q$ is
defined as
\begin{equation}
  \|P-Q\|_{\mathrm{TV}} = \frac{1}{2}\sum_x |P(x)-Q(x)|,
\end{equation}
and the Hellinger distance is defined as
\begin{equation}
  H(P,Q) = \frac{1}{\sqrt{2}}\sqrt{\sum_x \left( \sqrt{P(x)} - \sqrt{Q(x)}\right)^2 }.
\end{equation}
Both are metrics on the probability simplex, bounded between $0$ and $1$. To
estimate the convergence time, at each step we compute the distance (in both
metrics) between the distribution over the graph vertices and the steady-state
distribution, and record the first step at which it falls below a threshold of
$10^{-3}$. We simulate the number of steps required to reach this
threshold for fixed $A=100$ and $p=0.9$, varying the number of qubits in the
range $N\in[5,70]$. The results are shown in Fig.~\ref{fig:metrics_0}. For
small values of $N$, the convergence time initially increases and reaches a
maximum around $N\simeq 20$. For larger $N$ within the simulated range, the
convergence time decreases approximately linearly, with coefficient of
determination $R^2\simeq 0.99$ for a linear fit.

\begin{figure}[htbp]
  \centering
  \begin{subfigure}{\linewidth}
    \centering
    \includegraphics[width=\linewidth]{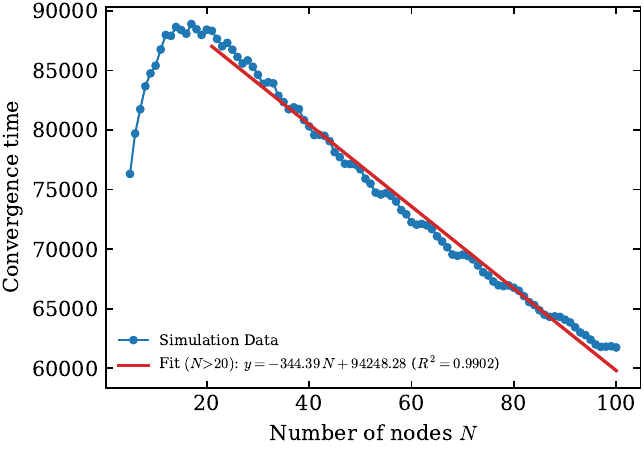}
    \caption{Convergence time in the total variation metric as a function of the number $N$ of nodes in the OQW for $A = 200$, $p = 0.7$.}
    \label{fig:metrics_1_tv}
  \end{subfigure}

  \medskip

  \begin{subfigure}{\linewidth}
    \centering
    \includegraphics[width=\linewidth]{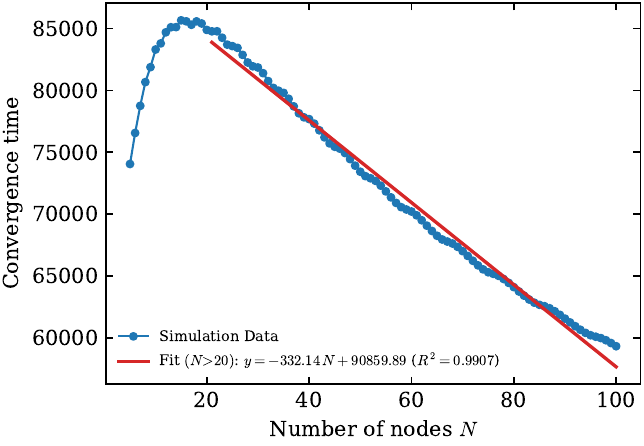}
    \caption{Convergence time in the Hellinger metric as a function of the number $N$ of nodes in the OQW for $A = 200$, $p = 0.7$.}
    \label{fig:metrics_1_hell}
  \end{subfigure}

  \caption{Total variation and Hellinger metrics to determine the convergence time as a function of the number $N$ of nodes in the OQW for $A = 200$, $p = 0.7$.}
  \label{fig:metrics_1}
\end{figure}

We repeat the same analysis for $A=200$ and $p=0.7$, obtaining the same
qualitative behavior, as shown in Fig.~\ref{fig:metrics_1}. This decrease
should not be interpreted as an asymptotic scaling law, since the positivity
condition imposes
\begin{equation}
N \leq \frac{2A}{\sqrt{p}}-1 .
\end{equation}
Thus, for fixed $A$ and $p$, the observed linear decay occurs only over the
finite admissible interval of $N$.

It is instructive to compare this numerical behavior with the analytic
cycle time of Eq.~\eqref{eq:n1exact}. At fixed $A$ and $p$ one has
\begin{equation}
  n_{1} = \frac{2A^{2}H_{N}}{p\,(N+1)} \;\sim\; \frac{2A^{2}}{p}\,\frac{\ln N}{N},
  \qquad N \gg 1,
\end{equation}
so the characteristic timescale of the walk decreases with $N$
throughout the admissible window $N \le 2A/\sqrt{p} - 1$. This is the
same statement as the growth of the mean drift velocity,
$v_{\mathrm{mean}} = p(N+1)(N+2)/6A^{2} \sim pN^{2}/6A^{2}$: at fixed
$A$, adding qubits accelerates the collective decay channel. We stress that the quadratic growth of
$v_{\mathrm{mean}}$ holds at fixed $A$; along the positivity boundary
$A^{2} = p(N+1)^{2}/4$ one finds instead
$v_{\mathrm{mean}} \to 2/3$ as $N \to \infty$.

Note, however, that the numerically observed convergence time is not governed by the mean cycle time
$n_1$. Indeed $n_1 = 2A^2 H_N / [p(N+1)]$ decreases monotonically with $N$,
whereas the data in Figs.~\ref{fig:metrics_0} and~\ref{fig:metrics_1} are manifestly
non-monotonic. The mean first-return time and the relaxation time of the
distribution are distinct quantities, and here they behave differently. The
relevant timescale is instead set by the spectral gap of the transition matrix
$T$ of Eq.~\eqref{eq:T}, which can be obtained in closed form.

Since $T$ is cyclic bidiagonal, 
$\det(T - \lambda I) = 0$ gives us the characteristic equation
\begin{equation}
  \prod_{k=0}^{N-1} \bigl( \lambda - q_k \bigr)
  \;=\; \prod_{k=0}^{N-1} p_k .
  \label{eq:chareq}
\end{equation}
Writing
$\lambda = 1 - \epsilon$ we obtain
$\prod_k (1 - \epsilon/p_k) = 1$, equivalently
$\sum_k \ln(1 - \epsilon/p_k) = 2\pi i m$ with $m = 0, \dots, N-1$ labelling the
eigenvalues. The positivity condition~\eqref{eq:Asuff} guarantees only
$p_k \le 1$; when $A$ lies well above the threshold $\tfrac{\sqrt p}{2}(N+1)$ --
as in the regimes of Figs.~\ref{fig:metrics_0} and~\ref{fig:metrics_1}, where
$\max_k p_k \lesssim 0.11$ -- one has $p_k \ll 1$ and may expand to second order.
Defining the moments
$S_j = \sum_{k=0}^{N-1} p_k^{-j}$, so that $S_1 = n_1$, this yields
\begin{equation}
  \epsilon_m \;\simeq\; \frac{2\pi^2 m^2 S_2}{S_1^3}
                 \;-\; i\,\frac{2\pi m}{S_1}.
  \label{eq:epsm}
\end{equation}
The subdominant modes are complex: they rotate around the ring with period
$n_1/m$, one lap of the walker, while decaying. The slowest is $m = 1$, so
\begin{equation}
  \Delta \;\equiv\; 1 - |\lambda_1|
  \;\simeq\; \frac{2\pi^2 (S_2 - S_1)}{S_1^3}
  \;\simeq\; \frac{2\pi^2 S_2}{S_1^3},
  \label{eq:gap}
\end{equation}
the last step because $p_k^{-2} \gg p_k^{-1}$. Evaluating $S_1$ and $S_2$ by the
same partial fraction used in Appendix~\ref{app:steady},
\begin{equation}
  S_1 = \frac{A^2}{p}\,\frac{2H_N}{N+1},
  \qquad
  S_2 = \frac{A^4}{p^2}\,
        \frac{2H_N^{(2)} + \dfrac{4H_N}{N+1}}{(N+1)^2},
  \label{eq:S1S2}
\end{equation}
with $H_N^{(2)} = \sum_{l=1}^{N} l^{-2} \to \pi^2/6$, we obtain the closed form
\begin{equation}
  \Delta \;\simeq\;
  \frac{\pi^2 p \bigl[\, 2(N+1) H_N^{(2)} + 4 H_N \,\bigr]}{4 A^2 H_N^3}
  \;\xrightarrow[N \gg 1]{}\;
  \frac{\pi^4\, p\, (N+1)}{12\, A^2\, H_N^3}.
  \label{eq:gapclosed}
\end{equation}
Since the total variation distance decays as $|\lambda_1|^t$, the convergence
time to a threshold $\varepsilon$ satisfies
$t_{\mathrm{conv}} \simeq \ln(C/\varepsilon)/\mathrm{gap}$, that is
\begin{equation}
  t_{\mathrm{conv}} \;\simeq\;
  \ln\!\Bigl(\frac{C}{\varepsilon}\Bigr)\,
  \frac{12\, A^2\, H_N^3}{\pi^4\, p\, (N+1)} .
  \label{eq:tconv}
\end{equation}

Equation~\eqref{eq:gapclosed} is a second-order estimate in $p_k$ and agrees with
exact numerical diagonalisation of $T$ to a few percent over the figure range:
the relative error is smallest, $\approx 2.5\%$, near $N \simeq 20$, and grows to
$\approx 14\%$ at $N = 5$ and again as $A$ approaches the positivity threshold,
where $\max_k p_k \to 1$ and the expansion is least accurate. Equation~\eqref{eq:tconv}
reproduces the observed non-monotonicity: the controlling factor
$H_N^3 / \bigl[\, 2(N+1) H_N^{(2)} + 4 H_N \,\bigr]$ rises while the numerator
$H_N^3 \sim \ln^3 N$ dominates and falls once $H_N^{(2)}$ saturates at $\pi^2/6$
and the denominator grows linearly, producing a broad maximum near $N \simeq 17$
(flat to within $0.5\%$ for $15 \le N \le 20$). This matches the peak of the
simulated convergence time, which also occurs at $N = 17$ in
Figs.~\ref{fig:metrics_0} and~\ref{fig:metrics_1}; the large-$N$ surrogate
$H_N^3/(N+1)$, by contrast, peaks earlier, near $N \simeq 12$. The prefactor
$A^2/p$ accounts for the rescaling between the two parameter sets. Fitting
$t_{\mathrm{conv}}$ against $1/\mathrm{gap}$ for the data of
Fig.~\ref{fig:metrics_0} gives a straight line through the origin of slope
$6.51 \pm 0.03$ ($R^2 = 0.997$). Identifying this slope with
$\ln(C/\varepsilon)$ in Eq.~\eqref{eq:tconv} and using $\varepsilon = 10^{-3}$
gives $C = \varepsilon\, e^{6.51} \approx 0.67$, i.e. an offset
$\ln C \approx -0.40$ below the ideal value $\ln(1/\varepsilon) = 6.91$ that $C=1$
would produce.

Three caveats are in order. The expansion leading to Eq.~\eqref{eq:epsm} assumes
$p_k \ll 1$ and degrades when $A$ approaches the positivity threshold
$\tfrac{\sqrt p}{2}(N+1)$, where $\max_k p_k \to 1$. The constant $C$ depends on
the initial distribution and is not predicted by the gap alone. Finally, because
the subdominant eigenvalues are complex, the approach to stationarity is
oscillatory with period $n_1$, so $t_{\mathrm{conv}}$ is not a perfectly smooth
function of $N$.

\subsection{Locality and physical cost of the Kraus operators}
\label{sec:locality}

The protocols above achieve their economy of steps at the price of operator
nonlocality, and it is important to state this plainly. In the GHZ construction the
Kraus operators $K_\pm = (I \pm X^{\otimes N})/2$ are the spectral
projectors of an $N$-body stabilizer; realising them as a physical channel is a
joint operation of all $N$ qubits. The Dicke construction is nonlocal for the same
reason: the jump operator $J_c$ is collective and the reset edge carries
$X^{\otimes N}$. The no-jump operator $K_c = \sqrt{I - J_c^\dagger J_c}$ makes this
concrete. $J_c^\dagger J_c$ contains two-body terms that transfer an excitation from one qubit to another, and
taking its square root couples all $N$ qubits inseparably. Hence $K_c$, though
diagonal on the symmetric subspace, cannot be written as a product of local
operations.

Consequently, the word ``efficient'' as used here refers to the number of walk
steps, and not to the hardware cost of implementing a single step. This should be
contrasted with the dissipative-engineering literature, where the physical content
of the schemes rests on the dissipators being quasi-local: the universal
dissipative computation of Ref.~\cite{Cirac09} and the Markovian entangled-state
preparation of Ref.~\cite{Kraus2008} both engineer few-body Lindblad operators, and
the cavity scheme of Ref.~\cite{Sweke2013} prepares $W$ states from local decay
channels assisted by a shared bosonic mode. Our Kraus operators are not of this
form.

The nonlocality of $J_c$ is not an artefact of the construction but a physical
property of the collective decay it models: a single bosonic mode coupled
symmetrically to $N$ emitters generates the collective lowering operator without
any $N$-body gate, and the stabilizer $X^{\otimes N}$ can be read out by a measurement of depth $O(N)$ using one ancilla. This does not make the
scheme local, but it indicates that the required resources are of a kind that
reservoir engineering and stabilizer measurement already supply.

\section{\label{sec:quantum-traj}Quantum Trajectories Method via Open Quantum Walks}

\par
\begin{figure}[]
\[\begin{tikzcd}
	{\ket{0}} &&&& {\ket{1}} \\
	\\
	&& {\ket{2}}
	\arrow["{K_0}", from=1-1, to=1-1, loop, in=60, out=120, distance=5mm]
	\arrow["{K_1}", curve={height=-6pt}, from=1-1, to=1-5]
	\arrow["{K_2}", curve={height=-6pt}, from=1-1, to=3-3]
	\arrow["{K_0}", curve={height=-6pt}, from=1-5, to=1-1]
	\arrow["{K_1}", from=1-5, to=1-5, loop, in=60, out=120, distance=5mm]
	\arrow["{K_2}", curve={height=-6pt}, from=1-5, to=3-3]
	\arrow["{K_0}", curve={height=-6pt}, from=3-3, to=1-1]
	\arrow["{K_1}", curve={height=-6pt}, from=3-3, to=1-5]
	\arrow["{K_2}"', from=3-3, to=3-3, loop, in=300, out=240, distance=5mm]
\end{tikzcd}\]
\caption{\label{fig:quantum_traj_oqw} OQW model implementing $m=3$ Lindblad operators for a given initial state.}
\end{figure}
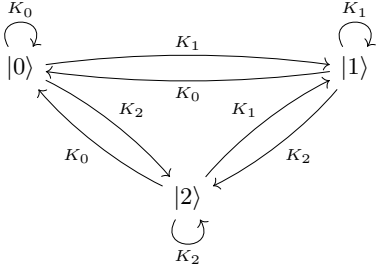

 To implement the quantum trajectories method within the OQW framework, we use the graph space as a classical register that records
the Kraus outcome associated with each step of the evolution. We describe the construction for $m$ Lindblad operators $\{L_j\}_{j=0}^{m-1}$ and an initial state $\rho_0 = \sum_{k=0}^{m-1} q_k \ket{\varphi_k}\bra{\varphi_k}$. We define the OQW via the Kraus operators
\begin{equation}
    M_i^j = K_j \otimes \ket{j}\bra{i},
\end{equation}
and the initial state
\begin{equation}
    \rho^{[0]} = \sum_k q_k \ket{\varphi_k}\bra{\varphi_k} \otimes \ket{k}\bra{k},
\end{equation}
where the operators $K_j$ are defined as the usual discretization of the Lindblad operators (see Appendix~\ref{app:quantum-traj}). As an example, for $m=3$ we obtain an OQW whose underlying graph is the complete graph on three vertices, as depicted in Fig.~\ref{fig:quantum_traj_oqw}.

\par After $n$ steps of the walk, we trace out the graph Hilbert space, obtaining
\begin{equation}
    \rho' = \sum_{i_1, \dots i_n} K_{i_n}\cdots K_{i_1}\, \rho_0 \,K_{i_1}^\dagger \cdots K_{i_n}^\dagger,
\end{equation}
which is precisely the quantum-trajectories evolution for $T = n\, d t$.

We emphasize that this construction is, in some sense, a reformulation of the quantum
trajectories method in the OQW language, rather than a claim of computational
speedup. Its purpose is to show that the stochastic structure of quantum
trajectories has a natural embedding in the OQW framework, where the graph
degree of freedom acts as a classical register for measurement outcomes. It is also important to emphasize that if $K_{0} = I - \tfrac{i}{\hbar} H_{\mathrm{eff}}\,dt$
and $K_{i} = \sqrt{dt}\,L_{i}$ one finds
$\sum_{i} K_{i}^{\dagger}K_{i} = I + \mathcal{O}(dt^{2})$.
The Kraus family is therefore trace preserving only to first order in
$dt$, as is standard for this discretization of the Lindblad generator;
the OQW completeness relation holds in the same sense.

\par This construction also connects to the framework of quantum collision models,
or repeated-interaction schemes, which trace back to Rau~\cite{Rau1963} and are now
a standard tool for open quantum dynamics~\cite{Ciccarello2022, Campbell2021}. There
the system interacts sequentially with a stream of ancillas, each collision
implementing one Kraus map. The quantum-trajectories walk above is of this type: the
transition operators $M_i^{\,j} = K_j \otimes \ket{j}\bra{i}$ apply the same Kraus
set $\{K_j\}$ at every vertex, and tracing out the graph returns the unconditional
evolution $\sum_j K_j(\cdot)K_j^\dagger$ obtained above. The underlying complete
graph of Fig.~\ref{fig:quantum_traj_oqw} reflects this: any outcome may follow any
other, so every ordered pair of vertices is an edge. The state-preparation protocols
of Secs.~\ref{sec:ghzstab} and~\ref{sec:dicke_oqw} use vertex-dependent transition
operators instead, and so are not of this collision-model form.	

\section{\label{sec:conclusion}Conclusion}

We have shown how open quantum walks can be used as a framework for
measurement-conditioned quantum state preparation with Kraus operators that are
not necessarily proportional to unitaries. In particular, we presented an OQW
construction for GHZ-type states based on projective measurements of the global
operator $X^{\otimes N}$, and a ring-shaped OQW protocol for preparing ensembles
of Dicke states. Since the W state is the $k=1$ Dicke state, the same
Dicke-state protocol also covers the preparation of W states. In the GHZ case,
the preparation is conditional on the walker measurement outcome and may require
a single-qubit phase correction. In the Dicke case, measuring the graph register
selects the desired Dicke component.

We also studied the convergence time for the Dicke-state protocol. Within the
admissible interval fixed by the positivity condition on the no-jump operator, which is the only regime in which the Kraus operators are well defined, the convergence
time decays approximately linearly with the number of nodes $N$ at fixed $A$ and
$p$, and we derived a closed-form spectral gap that reproduces this behavior,
including the non-monotonic onset at small $N$.

Extending the framework to other relevant classes of states, such as coherent
states \cite{Glauber1963} and cluster states \cite{Scarani2005}, remains open.
These states play central roles in quantum information processing
\cite{Menicucci2006, Raussendorf2001, Ralph2003}, and it would be interesting to
understand whether they admit natural OQW-based preparation protocols.

Finally, although the present work is theoretical, the Dicke-state construction
is motivated by experimentally well-established collective decay phenomena,
such as superradiance \cite{rosario25, Wang26}. This suggests that OQW-based
state-preparation protocols may provide a useful language for connecting
reservoir-engineering ideas with graph-based models of dissipative quantum
computation.

\section*{Data availability}
The data and code supporting the findings of this study are publicly available in the GitHub repository \texttt{state-prep-oqw} \cite{linckOQWStateprepRepo}.

\section*{Acknowledgments}
This study was financed in part by the Coordenação de Aperfeiçoamento de Pessoal de Nível Superior – Brazil (CAPES) – Finance Code 001. N.K.B. acknowledges financial support from CNPq Brazil (442429/2023-1) and FAPESP (Grant 2021/06035-0). The authors acknowledge the financial support of the National Institute of Science and Technology for Applied Quantum Computing through CNPq process No. 408884/2024-0.

\appendix

\section{\label{app:dicke-relations}Dicke state relations}
Here, we demonstrate the relations claimed in Sec.~\ref{sec:dicke_oqw} for Dicke states. First of all we have that
\begin{equation}
    \begin{split}
        J_c \ket{\mathcal{D}^N_k} & = \dfrac{\sqrt{p}}{\sqrt{\binom{N}{k}} A} \sum_i \sum_{x: |x| = k} J_i \ket{x_1\dots x_N}\\
        & = \dfrac{\sqrt{p}}{\sqrt{\binom{N}{k}} A} (N-k+1)\sum_{x : |x| = k-1} \ket{x_1\dots x_N} \\
        & = \dfrac{(N-k+1)\sqrt{p}}{\sqrt{\binom{N}{k}} A} \cdot \sqrt{\binom{N}{k-1}}\ket{\mathcal{D}^N_{k-1}} \\
        & = \sqrt{\frac{k p (N-k+1)}{A^2}}\, \ket{\mathcal{D}_{k-1}^N},
    \end{split}
\end{equation}
where $x\in\{0,1\}^N$ denotes a bit string and $|x|$ is its Hamming weight (i.e., the number of $1$'s in the string).
The first equality follows from the definition of $\ket{D_k^N}$, and the second equality follows from the fact that $J_i$
lowers the Hamming weight of $\ket{x_1\dots x_N}$ by one whenever it acts nontrivially. Moreover, each string of Hamming weight
$k-1$ has $N-(k-1)=N-k+1$ zeros, and each such zero can be obtained by lowering a string of Hamming weight $k$, obtained by
flipping exactly that zero into a one. For this combinatorial reason, the factor $N-k+1$ appears in the equation. The remaining steps follow by straightforward algebra.

\par By the same argument, we also have
\begin{equation}
J_c^\dagger \ket{\mathcal{D}_k^N}
= \frac{\sqrt{p}}{A}\sqrt{(k+1)(N-k)}\,\ket{\mathcal{D}_{k+1}^N}.
\end{equation}
Combining both relations, it is straightforward to see that
\begin{equation}
K_c \ket{\mathcal{D}_k^N}
= \sqrt{1 - \frac{k\,p\,(N-k+1)}{A^2}}\, \ket{\mathcal{D}_k^N},
\end{equation}
which proves the fundamental relations on which our results are based.

\section{\label{app:steady}Steady-state calculation}
We begin by computing the steady state of the discrete Markov chain
\cite{MC,SP, HSM} on the directed cycle
$\mathbb{Z}_{N}$, whose transition matrix is
 
\begin{equation}
  T =
  \begin{pmatrix}
    q_{0}   & 0       & \cdots  & 0       & p_{N-1} \\
    p_{0}   & q_{1}   & 0       & \cdots  & 0       \\
    0       & p_{1}   & q_{2}   & \ddots  & \vdots  \\
    \vdots  & \ddots  & \ddots  & \ddots  & 0       \\
    0       & \cdots  & 0       & p_{N-2} & q_{N-1}
  \end{pmatrix},
  \qquad q_{k} = 1 - p_{k},
  \label{eq:T}
\end{equation}
 
with $p_{k} = p(N-k)(k+1)/A^{2}$ the probability of hopping
$k \to k+1$ (indices mod $N$) and $q_{k}$ the probability of staying.
Because every node has exactly one outgoing hop, stationarity
$\pi = T\pi$ reduces to the statement that the probability current
$J_{k} \equiv \pi_{k} p_{k}$ is the same across every edge,
$\pi_{k} p_{k} = \pi_{k-1} p_{k-1} = J$ for all $k$. Hence
$\pi_{k} = J / p_{k}$, i.e.\ $\pi_{k} \propto 1/p_{k}$, and $J$ is fixed
by normalization. Thus
\begin{equation}\label{eq:sumB2}
    \begin{split}
        \sum_k 1/p_k & = \dfrac{A^2}{p}\sum_{k < N} \dfrac{1}{(N-k)(k+1)} \\
        & = \dfrac{A^2}{p} \sum_{k < N}\dfrac{1}{N+1} \left( \dfrac{1}{N-k} + \dfrac{1}{k+1}\right) \\
        & = \dfrac{A^2}{p}\dfrac{2H_N}{N+1},
    \end{split}
\end{equation}
which yields Eq.~(41), with $H_{N} = \sum_{l=1}^{N} 1/l$ the $N$-th
harmonic number. Note that Eq.~\eqref{eq:sumB2} is also, directly, the
expected cycle time $n_{1}$ of Eq.~\eqref{eq:n1exact}.

\section{\label{app:quantum-traj}Quantum Trajectories}
The quantum trajectories method is an established technique \cite{Daley14} for simulating quantum dynamics described by a Lindblad equation,
\begin{equation}
\dfrac{d}{dt}\rho(t) = \dfrac{1}{i\hbar}[H,\rho(t)] + \sum_j \left( L_j \rho(t) L_j^\dagger -\dfrac{1}{2}\left\{L_j^\dagger L_j, \rho(t) \right\} \right),
\end{equation}
where $L_j$ denotes the jump operators, also known as Lindblad operators. We define the effective Hamiltonian as
\begin{equation}
    H_{\text{eff}} = H - \dfrac{i \hbar}{2} \sum_j L_j^\dagger L_j .
\end{equation}
We can express the Lindblad equation as
\begin{equation}
\dfrac{d}{dt}\rho(t) = \dfrac{1}{i \hbar}\bigl(H_{\text{eff}}\rho(t) - \rho(t) H_{\text{eff}}^\dagger\bigr) + \sum_j L_j \rho(t) L_j^\dagger.
\end{equation}
Given a pure state $\ket{\varphi(t)}$, we also define
\begin{equation}
    \ket{\varphi_{\text{eff}}(t+dt)} = \left(1- \dfrac{i}{\hbar} H_{\text{eff}}\, d t\right)\ket{\varphi(t)}.
\end{equation}
Note that this is generally not a normalized state, since
\begin{equation}
\begin{split}
\langle \varphi_{\text{eff}}(t+d t) |\varphi_{\text{eff}}(t+d t) \rangle
& = 1 - \dfrac{i}{\hbar} d t \,\langle \varphi(t)|H_{\text{eff}} - H_{\text{eff}}^\dagger| \varphi(t) \rangle \\
& = 1 - \sum_j d t\,\langle \varphi(t)|L_j^\dagger L_j|\varphi(t) \rangle .
\end{split}
\end{equation}
We interpret $dp_j = d t\, \bra{\varphi(t)}L_j^\dagger L_j\ket{\varphi(t)}$ as the probability that the jump $L_j$ occurs within the time interval $d t$. Defining $dp = \sum_j dp_j$, the probability of the no-jump evolution generated by $H_{\text{eff}}$ in this infinitesimal interval is $1-dp$. This establishes the foundation for the quantum trajectories method discussed next.

First, we set the initial condition $\rho_0 = \sum_k q_k \ket{\varphi_k}\bra{\varphi_k}$. We aim to simulate the evolution of $\rho_0$ from $t = 0$ to $t = T$. We subdivide $T$ into $n$ steps of size $d t$. Next, we randomly sample $\ket{\varphi_k}$ with probability $q_k$ and then proceed with the evolution of $\ket{\varphi_k(t)}$ as follows:
\begin{enumerate}
    \item With probability $1-dp$ it evolves to $\ket{\varphi_k(t+dt)} = \dfrac{\ket{\varphi_{k,\text{eff}}(t+d t)}}{\sqrt{1-dp}}$;
    \item With probability $dp$ it evolves according to a jump operator. In this case, with probability $dp_m/dp$ it evolves to $\ket{\varphi_k(t+dt)} = \dfrac{L_m\ket{\varphi_k(t)}}{\sqrt{dp_m}}$;
    \item This procedure is repeated iteratively until we reach $\ket{\varphi_k(T)}$; we then store the result and restart the sampling with a new state $\ket{\varphi_\ell}$ in order to average the outcomes over many trajectories.
\end{enumerate}
The connection to the Kraus-operator formalism is established by defining
\begin{equation}
\label{eqn:krauszero}
    K_0 = I + \left(L_0 - \dfrac{i}{\hbar}H \right)d t,
\end{equation}
\begin{equation}
\label{eqn:krausops}
    K_i = \sqrt{d t}\, L_i, \qquad i > 0.
\end{equation}
It is well known \cite{OQS} that
\begin{equation}
    L_0 = -\dfrac{1}{2} \sum_j L_j^\dagger L_j.
\end{equation}
Therefore, we can write
\begin{equation}
    K_0 = I - \dfrac{i}{\hbar} H_{\text{eff}}\, d t.
\end{equation}

\bibliographystyle{apsrev4-2}
\bibliography{references}

\end{document}